\documentstyle[12pt]{article}
\input{psfig}

\def\Mpc{\,{\rm Mpc}}

\def\cmm2{{\,\rm cm^{-2}}}
\def\cm2{{\,{\rm cm}^2}}
\def\cmm3{{\,{\rm cm}^{-3}}}
\def\gcmm3{{\,{\rm g\,cm^{-3}}}}
\def\kms{\,{\rm km\,s^{-1}}}

\def\fun#1#2{\lower3.6pt\vbox{\baselineskip0pt\lineskip.9pt
  \ialign{$\mathsurround=0pt#1\hfil##\hfil$\crcr#2\crcr\sim\crcr}}}

\begin{document}
\pagestyle{empty}
\begin{center}
\bigskip

\rightline{FERMILAB--Pub--96/359-A}
\rightline{astro-ph/9610158}
\rightline{to appear in {\em New Astronomy}}

\vspace{.2in}
{\Large \bf Big-bang Nucleosynthesis:\\
{\hfill} \\
Is the Glass Half Full or Half Empty?}
\bigskip

\vspace{.15in}
Michael S. Turner\\

\vspace{.1in}
{\it Departments of Astronomy \& Astrophysics and of Physics \\
Enrico Fermi Institute, The University of Chicago, Chicago, IL~~60637-1433}\\

\vspace{0.1in}
{\it NASA/Fermilab Astrophysics Center\\
Fermi National Accelerator Laboratory, Batavia, IL~~60510-0500}\\

\end{center}

\pagestyle{plain}
\setcounter{page}{1}
\vspace{0.2in}

Big-bang nucleosynthesis is a scientific success story
and a pillar of the standard hot big-bang cosmology.  Or is it?
Over the past year there has been a lively debate about just this.
As befits the times, the debate has been carried out on the
Los Alamos archive, at meetings and workshops, in coffee rooms,
and occasionally in refereed journals.  The paper by Fields, Kainulainen,
Olive and Thomas \cite{fieldsetal}
in the inaugural issue of {\em New Astronomy}
is part of this debate.

First some history; in the 1940s Gamow and his collaborators
put forth the idea that all the chemical elements could have been
synthesized a few minutes after the big bang, provided it was a
hot big bang.  (As it turns out, Coulomb barriers prevent significant
production of elements beyond mass eight.)  In 1948
Gamow's colleagues Alpher and
Herman used the yield of $^4$He and heavier elements
to predict the temperature of the Universe today
and arrived at 5\,K.  It seems that only Fred Hoyle
took this prediction seriously, and used it to argue against the big-bang
model:  A temperature of 5\,K exceeds the 2.3\,K upper limit
inferred from the relative abundance of rotationally excited CN molecules
in gas clouds in the Galaxy by Adams and McKellar in 1941.
As it turns out, the Alpher-Herman prediction was high because
of the value of the Hubble constant used and the assumption
that baryons provide the critical density, and the Adams-McKellar
temperature limit was low.

The rest of the story is well known.  In the early 1960s, unaware
of Gamow's work, Peebles repeated the calculations and convinced his Princeton
colleagues Dicke, Roll, and Wilkinson to search for this microwave
radiation.  Before they could, down the road in Holmdel, NJ
Penzias and Wilson
made their serendipitous discovery of the Cosmic Background Radiation (CBR).
The temperature of the CBR is now know to four significant
figures, $T=(2.728\pm
0.002)\,$K, thanks to the beautiful measurement by the FIRAS instrument
on the COBE satellite \cite{firas}.  Shortly after the discovery of the CBR,
the first detailed calculations of big-bang nucleosynthesis
were carried out by Peebles and by Wagoner, Fowler and Hoyle.

\begin{figure}[t]
\centerline{\psfig{figure=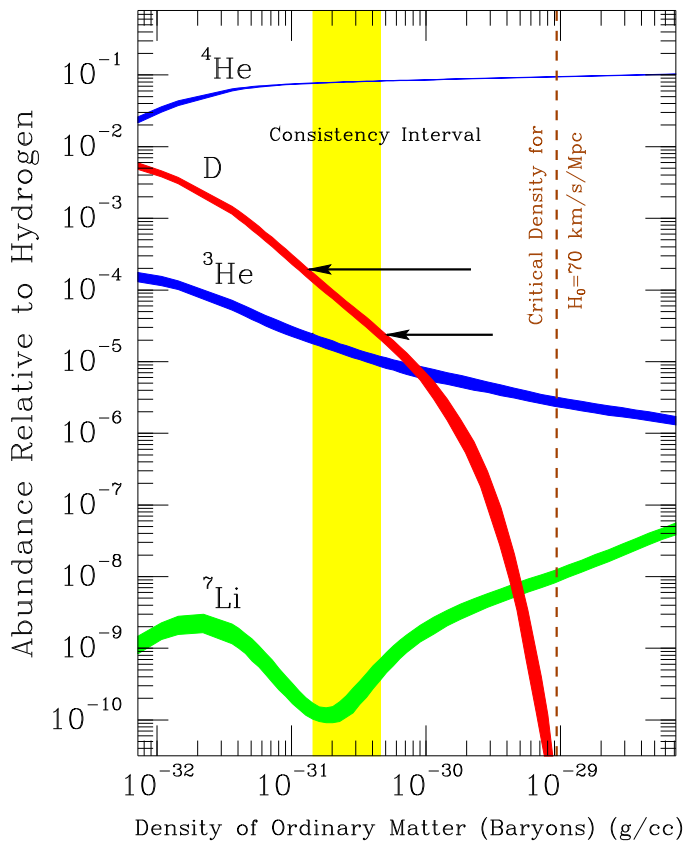,width=3.5in}}
\caption{Big-bang production of the light elements;
widths of the curves show the two-sigma theoretical uncertainty.
The pre-debate consensus consistency interval is shown
($\rho_B = 1.5\times 10^{-31}\gcmm3 - 4.5\times 10^{-31}\gcmm3$, or
$\Omega_Bh^2 = 0.008 - 0.024$).  Arrows indicate the high and low
deuterium detections.}
\end{figure}

Four light nuclei are produced in significant amounts
-- D, $^3$He, $^4$He and $^7$Li -- with
the yields that depend upon the baryon density and input microphysics
(nuclear cross sections and the number of light neutrino species).
The yield of $^4$He is large (by mass around 25\%) and varies
logarithmically with baryon density (Fig. 1).  Establishing
the existence of a large, primeval $^4$He abundance was the first
success of big-bang nucleosynthesis \cite{pp}.

The yield of deuterium is much smaller,
by number around $10^{-5}$ relative to H (Fig.~1); moreover,
the deuteron is weakly bound and easily destroyed.  However, in 1973
Reeves, Audouze, Fowler
and Schramm \cite{rafs} made the case for deuterium's cosmological
utility:  It cannot be made in significant amounts in the
contemporary Universe -- the mere presence of deuterium is evidence
for the big bang -- and the rapid decrease of its big-bang
production with baryon density makes it a good ``baryometer.''
The first determination of the
cosmic deuterium abundance, indirectly in
the solar wind by a foil placed on the moon by Apollo astronauts
and in the local ISM by the Copernicus satellite, was
the second success of big-bang nucleosynthesis.
Further, by setting a lower limit to the primeval D abundance,
it led to an upper limit to the baryon density,
at most 10\% of the critical density.
To foreshadow, an actual determination -- as opposed to a lower
limit -- of the primeval D abundance allows an accurate measure of
the baryon density.

The $^3$He and $^7$Li stories are more complicated; both are produced
and destroyed in the contemporary Universe.  The abundance of
$^7$Li varies from greater than $10^{-9}$ relative to H in meteorites
to less than $10^{-12}$ relative to H in some low-mass stars.  In the early 1980s
the Spites \cite{spites} announced that they
had determined the primeval $^7$Li abundance by measuring
its abundance in the atmospheres of pop II halo stars.
Their case hinged upon ``the Spite
plateau'' -- a leveling of the abundance with increasing stellar mass
at a value around $(1.5 \pm 0.5) \times 10^{-10}$ relative to H.
Lithium can be destroyed by convection; lower-mass stars have deeper convection zones; the leveling of
the $^7$Li abundance was indicative of
the disappearance of convective $^7$Li burning.  The abundance
measured by the Spites was consistent with the abundances of D and $^4$He;
success number three.

All stars produce $^3$He by burning D even before they reach the
main sequence; low-mass stars are believed
to make additional $^3$He and high-mass stars destroy most of their
$^3$He.  Since the material in the ISM is either primeval or cycled
through stars -- mainly through low-mass stars since metal production
by massive stars limits the amount of processing they can do --
it has been argued that the sum of D + $^3$He
has not changed greatly.  Based upon this argument, $^3$He was brought
into the fold in the early 1980s \cite{ytsso}; success number four.

Until a year ago most workers in the field
would have agreed that the big-bang predictions for all four light
nuclei are consistent with their measured abundances provided
the baryon density is between $1.5\times 10^{-31}\gcmm3$ and
$4.5\times 10^{-31}\gcmm3$ \cite{many}, corresponding to a fraction of critical
density $\Omega_B \simeq (0.01- 0.02)h^{-2}$ ($h=H_0/100\kms\Mpc^{-1}$).
This is the earliest test of the hot big bang and establishes
a firm foundation for the exciting speculations about the Universe
at even earlier times (e.g., inflation and cold dark matter).
Accepting the success of the standard cosmology, this leads to the best
determination of the density of ordinary matter as well as a
stringent limit to the number of light neutrino species, $N_\nu
< 4$ \cite{ssg} (a prediction now confirmed by the high-precision
LEP/SLC direct measurements based upon the width of the Z-boson).

The success as well as the importance of nucleosynthesis has spurred
increased interest and more observations.  The abundance of $^7$Li
(as well as $^6$Li, B and Be) has now been measured in
hundreds of old halo stars \cite{deplete}; there are now high-precision measurements
of the $^4$He abundance in more than fifty metal-poor, extragalactic
HII regions \cite{pagel}.  Within the past year the $^3$He abundance has been measured
in the local ISM for the first time \cite{gg}, HST has accurately
determined the D abundance in the local ISM \cite{linsky},
and a twenty-year old goal has
been realized -- detection of D in high redshift hydrogen clouds
($z\sim 2.5 - 4.7$) \cite{st}.  These new observations have made possible discussions
-- in some cases arguments -- about the third significant figure in
the primordial $^4$He abundance, about the extent to which $^7$Li may
have been depleted in old halo stars, about whether or not low-mass
stars preserve and produce additional $^3$He, and perhaps most interestingly
the value of the primeval D abundance.

There are three detections of the D Ly-$\alpha$ feature in the absorption
spectra of high-redshift QSOs -- two by Tytler and
his colleagues \cite{tytler} in clouds at redshifts $z=2.5$ and $z=3.57$
and one by Songaila and her colleagues \cite{songaila}
in a cloud at redshift $z=3.32$; there
are four other tentative detections.  Both of Tytler's clouds
give (D/H)$\simeq (2.4\pm 0.3)\times 10^{-5}$, while Songaila's cloud gives
a value that is about ten times larger,
(D/H)$\simeq (2\pm 0.4)\times 10^{-4}$.  The Tytler value is at the extreme
low end of the anticipated range, corresponding to the highest baryon
density; the Songaila value is at the extreme high end of the
anticipated range (in my book, this is success number five).
Since a measurement of the primeval deuterium
abundance pegs the baryon density very accurately
much of the recent debate centers on it.  Tytler et al suggest that
the Songaila detection is due to a rogue hydrogen cloud fortuitously
located to mimic D, while Songaila et al suggest that Tytler has underestimated
the neutral hydrogen column and/or deuterium has been depleted in his clouds.

Fields and his colleagues \cite{fieldsetal} favor the high value of deuterium,
which indicates a low baryon density $\rho_B \simeq 1.5\times
10^{-31}\gcmm3$ ($\Omega_B \simeq 0.01h^{-2}$), because the
predicted $^4$He and $^7$Li abundances then nicely fit
the observations.  This interpretation makes the
case for nonbaryonic dark matter ironclad since $\Omega_B$ can be
at most a few percent and few would argue that $\Omega_0$ can even
be as small as 10\%.  The problem is explaining where
all the deuterium went.  In the ISM today, (D/H)$\simeq (1.6\pm 0.1)
\times 10^{-5}$, about a factor of ten smaller.  Further, this, taken
with Gloeckler and Geiss's measurement of $^3$He in the local ISM
today \cite{gg}, implies (D+$^3$He)/H$\simeq (3.7\pm 0.8) \times 10^{-5}$,
a factor of five smaller than the primordial value.  To accommodate this
requires an efficient new way of destroying of $^3$He  -- several
have been suggested \cite{killhe3} -- but even that is not an easy out.  The
value of D + $^3$He deduced for the pre-solar nebula, which
reflects the ISM 4.5 Gyr ago, is essentially identical,
(D+$^3$He)/H$\simeq (4.2\pm 1) \times 10^{-5}$, suggesting
that an efficient mechanism of destroying $^3$He is not at work \cite{ttsc}.
In any case, $^3$He is a problem as the Gloeckler and Geiss's
measurement of $^3$He is not consistent with the standard picture that
the cosmic abundance of $^3$He slowly increases with time due to
production by low-mass stars.

Others, including Steigman and his colleagues \cite{others}, favor Tytler's
low value of primeval deuterium.  Then the baryon density is at the
high end, $\rho_B \simeq 4.5\times 10^{-31}\gcmm3$ or $\Omega_B
\simeq 0.02h^{-2}$ (the case for nonbaryonic dark matter is still
strong as $\Omega_B$ must still be less than 10\%).  Problems
with D and $^3$He disappear (the small D depletion from
the big bang until the present is a little puzzling).
However, now $^7$Li and $^4$He are problematic.  The Spite-plateau abundance
is only about half the big-bang prediction and the predicted
$^4$He abundance is $Y_P = 0.245$ compared to the most frequently
quoted analysis of the primordial abundance $Y_P=0.232\pm 0.003 ({\rm stat})
\pm 0.005 ({\rm sys})$ \cite{os}.  Accommodating this requires
$^7$Li depletion in old halo stars -- there is some observational
evidence for depletion \cite{deplete} and some theoretical models predict depletion
\cite{yale} -- and not taking
the errors on the $^4$He abundance at face value -- some have argued
that the systematic errors are a factor of two larger \cite{skillman}
and a new determination of the primeval $^4$He abundance based upon new objects is
higher by about 0.01 \cite{izotov}.  (Steigman and his colleagues
suggest a more radical solution \cite{hata}:  new physics in the form
of a 10-MeV tau neutrino, which would lead to a reduction
in the predicted $^4$He abundance.)

The quest to pin down the baryon density to 10\% and
sharply test the big bang with all four light elements
is on.  A flood of high-quality measurements of
light-element abundances -- from deuterium in high redshift
clouds to $^7$Li in halo stars -- is rolling in.  There is presently
some confusion, due to a poor understanding of
systematic errors ($^4$He and primeval deuterium) and uncertainty
about galactic chemical evolution ($^3$He) and stellar processing ($^3$He
and $^7$Li).  I am confident that theory aided by
additional observations (or vice versa) will sort things
out in the next few years.
About the time this happens, there will be a beautiful
independent test:  A precision determination of the baryon density from
the mapping of CBR anisotropy on very small scales (arcminutes to a degree).
These measurements will made by balloon-based bolometers,
ground-based interferometers and two new satellites
(NASA's MAP and ESA's COBRAS/SAMBA).
A comparison with the nucleosynthesis determination of the baryon density
will be a crucial test of the standard cosmology.

Is the glass half full or half empty?  My assessment is
half full; not all may agree, but I believe all agree
that this is an exciting time.

\end{document}